\documentclass[aps,prb, twocolumn,reprint,superscriptaddress]{revtex4-1}

\usepackage{graphicx}
\usepackage[colorlinks=true,urlcolor=blue]{hyperref}
\usepackage[applemac]{inputenc}
\usepackage[T1]{fontenc}

\begin{document}

\title{Subgap kinetic inductance detector sensitive to 85-GHz radiation}

\author{F.~Levy-Bertrand}
\email{florence.levy-bertrand@neel.cnrs.fr}
\affiliation{Univ. Grenoble Alpes, CNRS, Grenoble INP, Institut N\'eel, 38000 Grenoble, France}

\author{A.~Beno\^it}
\affiliation{Univ. Grenoble Alpes, CNRS, Grenoble INP, Institut N\'eel, 38000 Grenoble, France}

\author{O.~Bourrion}
\affiliation{Laboratoire de Physique Subatomique et de Cosmologie, Universit\'e Grenoble Alpes, CNRS, 53 rue des Martyrs, 38026 Grenoble Cedex, France}

\author{M.~Calvo}
\affiliation{Univ. Grenoble Alpes, CNRS, Grenoble INP, Institut N\'eel, 38000 Grenoble, France}

\author{A.~Catalano}
\affiliation{Laboratoire de Physique Subatomique et de Cosmologie, Universit\'e Grenoble Alpes, CNRS, 53 rue des Martyrs, 38026 Grenoble Cedex, France}

\author{J.~Goupy}
\affiliation{Univ. Grenoble Alpes, CNRS, Grenoble INP, Institut N\'eel, 38000 Grenoble, France}

\author{F.~Valenti}
\affiliation{IPE,~Karlsruhe~Institute~of~Technology,~76344~Eggenstein-Leopoldshafen,~Germany}
\affiliation{PHI,~Karlsruhe~Institute~of~Technology,~76131 Karlsruhe,~Germany}

\author{N.~Maleeva}
\affiliation{PHI,~Karlsruhe~Institute~of~Technology,~76131 Karlsruhe,~Germany}
\affiliation{National University of Science and Technology MISIS, 119049 Moscow, Russia}

\author{L. Gr\"unhaupt}
\affiliation{PHI,~Karlsruhe~Institute~of~Technology,~76131 Karlsruhe,~Germany}

\author{I.~M.~Pop}
\affiliation{IQMT,~Karlsruhe~Institute~of~Technology,~76344~Eggenstein-Leopoldshafen,~Germany}
\affiliation{PHI,~Karlsruhe~Institute~of~Technology,~76131 Karlsruhe,~Germany}

\author{A.~Monfardini}
\email{alessandro.monfardini@neel.cnrs.fr}
\affiliation{Univ. Grenoble Alpes, CNRS, Grenoble INP, Institut N\'eel, 38000 Grenoble, France}

\begin{abstract}
We have fabricated an array of subgap kinetic inductance detectors (SKIDs) made of granular aluminum ($T_c\sim$2~K) sensitive in the 80-90 GHz frequency band and operating at 300~mK. We measure a noise equivalent power of $1.3\times10^{-16}$~W/Hz$^{0.5}$ on average  and $2.6\times10^{-17}$~W/Hz$^{0.5}$ at best, for an illuminating power of 50~fW per pixel. Even though the circuit design of SKIDs is identical to that of the kinetic inductance detectors (KIDs), the SKIDs operating principle is based on their sensitivity to subgap excitations. This detection scheme is advantageous because it avoids having to lower the operating temperature proportionally to the lowest detectable frequency. The SKIDs presented here are intrinsically selecting the 80-90 GHz frequency band, well below the superconducting spectral gap of the film, at approximately 180 GHz.

\vspace{0.3cm}
\noindent To cite: 

\noindent  F. Levy-Bertrand et al, Physical Review Applied 15, 044002 (2021). 

\noindent \href{http://dx.doi.org/10.1103/PhysRevApplied.15.044002}{DOI:} 10.1103/PhysRevApplied.15.044002
\end{abstract}

\maketitle

\section{Introduction}

Terahertz (THz) electromagnetic waves are defined from 0.1~THz to 10~THz, corresponding to wavelengths from 3~mm down to 30~$\mu$m~\cite{tonouchi_cutting-edge_2007,rogalski_terahertz_2011,dhillon_2017_2017}. These waves are nonionizing and can pass through wood, plastic, clothing and  the atmosphere in a few frequency bands (THz atmospheric transmission windows). Most molecules have vibration or rotation frequencies in the THz range, allowing spectroscopic identification~\cite{encrenaz_detectability_1995,phillips_submillimeter_1992}. These properties drive the development of THz imaging and THz spectroscopy for various potential applications such as security~\cite{mittleman_sensing_2013}, medical diagnostic~\cite{siegel_terahertz_2004}, quality control processes and astrophysics research~\cite{monfardini_latest_2014,schlaerth_millimeter_2008}. To better exploit these potential applications, two main research directions are being conducted: the development of sensitive detectors and the development of intense sources. The present work focuses on the investigation of a sensitive detector, the subgap kinetic inductance detector~\cite{dupre_tunable_2017} (SKID), a particular implementation of the kinetic inductance detector concept (KID)~\cite{day_broadband_2003,doyle_lumped_2008,mccarrick_horn-coupled_2014}. 

KIDs are detectors used for millimeter-wave observations in astrophysics~\cite{patel_fabrication_2013,matsumura_mission_2014,griffin_spacekids_2016,baselmans_performance_2016} and for passive THz cameras emerging for security applications~\cite{rowe_passive_2016,noauthor_httpssequestimcomtechnology_nodate}. They are planar resonant circuits made of superconductors deposited on an insulating substrate. They are usually cooled down to about 100~mK.  The photon detection principe consists of monitoring the resonance frequency shift that is proportional to the incident power. Thanks to a straightforward fabrication process, an easy multiplexing technique, and a compact design, KIDs are ideal for achieving large sensitive arrays with thousands of pixels, such as those currently installed in the NIKA2 millimetric camera of the IRAM 30~m telescope (Pico Veleta, Spain)~\cite{adam_nika2_2018}. The current NIKA2 KID arrays have a total of about 3000 KIDs fabricated from thin superconducting aluminum, operating at 150~mK, and detecting frequencies in the 120-300~GHz band. Their noise equivalent power is of the order of 10$^{-17}$W/Hz$^{0.5}$ to 10$^{-16}$W/Hz$^{0.5}$ as required for ground-based observations.

The circuit design of SKIDs is identical to that of KIDs. The photon detection principe also consists of monitoring the resonance frequency shift that is proportional to the incident power. However, the mechanism giving rise to the frequency shift is different. For classic KIDs~\cite{day_broadband_2003}, in order to generate a frequency shift the incident photon must carry an energy $h\nu$ higher than or equal to the superconducting spectroscopic gap $2\Delta$. This gap is the minimum energy required to break the Cooper pairs forming the superfluid. Thus, for a given superconducting material, the smallest detectable frequency is $\nu_{low}=2\Delta/h$. An associated constraint is that, in order to exclude temperature-induced excitations, the operating temperature must be less than or equal to $T\sim T_c/10\sim2\Delta/(35k_B)$ where $T_c$ is the superconducting critical temperature and $k_B$ is the Boltzmann constant. In short, for classic KIDs, reducing the smallest detectable frequency requires a proportional reduction of the superconducting gap, and, consequently, of the operating temperature. For detection at 80~GHz, an operating temperature lower than $\sim$100~mK would thus be needed. The working principle of SKIDs removes this constraint. For SKIDs~\cite{dupre_tunable_2017,levy-bertrand_electrodynamics_2019}, the shift of the measured frequency is due to absorbed photons at a subgap excitation. The frequency shift occurs because in superconducting thin films, the superfluid density, and thus the kinetic inductance, depends on the circulating current~\cite{gennes_superconductivity_nodate, swenson_operation_2013, dupre_tunable_2017}. The current flowing in the detector is increased by the absorbed photons.
For a current $I$ that is small compared with the critical current $I_c$, the frequency shift is as follows~\cite{gennes_superconductivity_nodate, kher_kinetic_2016}: 
\begin{eqnarray}
\frac{\delta f}{f}\sim-\frac{\alpha}{2}\big(\frac{I}{I*}\big)^2
\label{eq_sensitivity}
\end{eqnarray}
where $\alpha$ is the ratio of the kinetic inductance by the total inductance and $I*=3\sqrt{3}/2\times I_c$.
To detect small incident power, equation~(\ref{eq_sensitivity}) provides evidence that a small critical current $I_c$ is needed. With this aim, two parameters can be optimized: the critical current density $J_c$ , that is material dependent, and the section of the kinetic inductance design. This effect can also be formalized in the language of circuit quantum electrodynamics, using the cross-Kerr coefficient~\cite{maleeva_circuit_2018}.

\begin{figure}
\begin{center}
\resizebox{8cm}{!}{\includegraphics{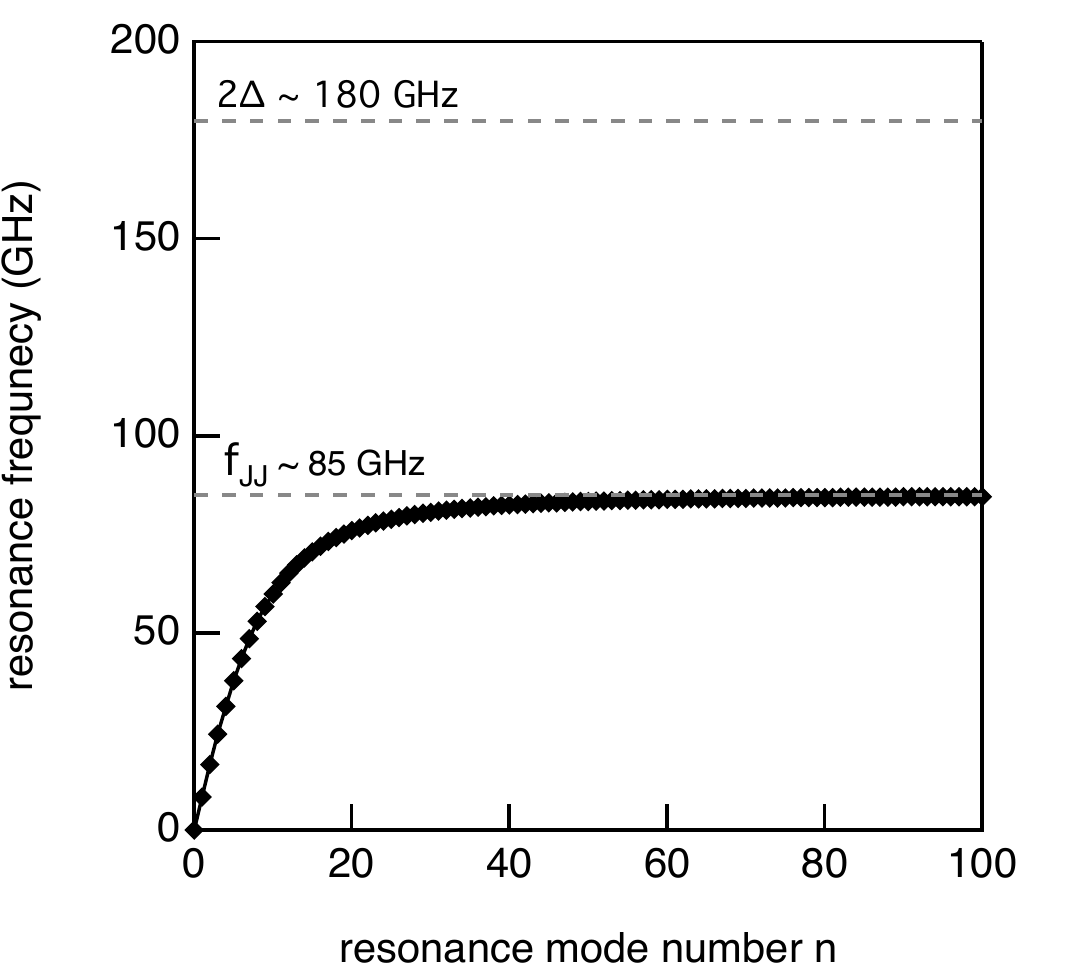}}
\caption{\textbf{Dispersion relation of the SKIDs under study.} The model used to compute the dispersion relation assumes a resonator made of a superconducting material  consisting of an array of Josephson junctions~\cite{maleeva_circuit_2018}. The dispersion relation saturates at the  Josephson junction plasma frequency  $f_{JJ}$.}
\label{fig_Wjj}
\end{center}
\end{figure}

Subgap excitation frequencies are naturally present in most superconducting resonators. Indeed, in addition to the fundamental mode, a superconducting resonator can usually support several higher harmonics with frequencies below the superconducting spectroscopic gap. For the lumped resonator design employed for the SKIDs, the fundamental resonance $f_1$ is of the order of 3~GHz, and, although the harmonic resonance frequencies $f_n$ are not exactly multiples, they can be roughly estimated by $f_n=n\times f_1$ where $n$ is an integer indexing the resonance mode. This linear relation remains valid almost all the way up to the surface plasma frequency where the dispersion relation saturates. For a low kinetic inductance superconductor (such as thin film Al, Nb, Ta, etc.) the  surface plasma frequency is much larger than the superconducting spectroscopic gap. However, for a superconducting material  consisting of an array of Josephson junctions, the dispersion relation saturates at the Josephson junction plasma frequency $f_{JJ}$, that can be smaller than the superconducting spectroscopic gap.  Maleeva and coauthors model the granular aluminum (grAl) thin films as a network of Josephson junctions~\cite{maleeva_circuit_2018}. In the model, the Josephson junction plasma frequency is given by
\begin{eqnarray}
f_{JJ}=\frac{1}{2\pi}\sqrt{\frac{2eI_c}{\hbar C_J}}
\label{eq_wjj}
\end{eqnarray}
where $C_J$ is the effective capacity between the superconducting grains. Note that the grains in the model do not necessarily correspond to the actual grains in the film, because several grains can be strongly coupled together and collectively charge as one object, as recently evidenced by scanning tunneling experiments~\cite{yang_microscopic_2020}. Figure~\ref{fig_Wjj} shows a typical dispersion relation for a critical current density $J_c~\sim$~4~mA/$\mu$m$^2$ and $C_J~\sim$~40~fF/$\mu$m$^2$. These values are characteristics of a granular aluminum with a normal state resistivity of $\sim$~900~$\mu\Omega$.cm~\cite{maleeva_circuit_2018}. As the density of subgap modes diverges at the plasma frequency, a strong photon absorption is expected.

In theory, the frequency selectivity is adjusted via the Josephson junction plasma frequency. In practice, a variation of the Josephson junction plasma frequency with the normal state resistivity of granular aluminum is observed~\cite{levy-bertrand_electrodynamics_2019}. Table~\ref{table} lists, from this previous study~\cite{levy-bertrand_electrodynamics_2019}, the variation of the Josephson junction plasma frequency with the normal state resistivity. Owing to the fact that the granular aluminum film thickness is not much larger than the effective grain size, possibly giving rise to surface charging effects, and the fact that the effective intergrain capacitance likely has a nontrivial dependence on the microstructure of the film and its resistivity, the quantitative modeling of the plasma frequency as a function of thickness and resistivity is still an open question.  

\begin {table}[h]
\caption {\textbf{Measured Josephson junction plasma frequency for  granular aluminum films with different normal state resistivity}, where $\rho$ is the normal state resistivity, $d$ is the film thickness, $\Delta$ is the superconducting gap, and $f_{JJ}^{exp}$ is a strong subgap absorption interpreted as a Josephson junction plasma frequency~\cite{levy-bertrand_electrodynamics_2019}. } 
\label{table}
\begin{center}
\begin{tabular}{cccc}
\hline
\hline
$\rho~(\mu \Omega$cm)& $d$ (nm) & $2\Delta/h$ (GHz) & $f_{JJ}^{exp}$ (GHz)\\
\hline
900&20&181&84\\
1600&20&182&81\\
2000&20&179&78\\
3000&30&165&73\\
\hline
\hline
\label{table}
\end{tabular}
\end{center}
\end {table}

Here we report the performances of a  SKID array made of superconducting granular aluminum with  a normal state resistivity of $\sim$~900~$\mu\Omega$.cm. The intense sensitivity at about  85~GHz is interpreted as a divergence of the density of subgap modes at the plasma frequency~\cite{maleeva_circuit_2018} (see Fig.~\ref{fig_Wjj}), explaining why the SKIDs are intrinsically selecting the 80-90 GHz frequency band. These SKIDs may provide a scalable technology for imaging at 85~GHz.

\section{Experimental}

\begin{figure}
\begin{center}
\resizebox{8cm}{!}{\includegraphics{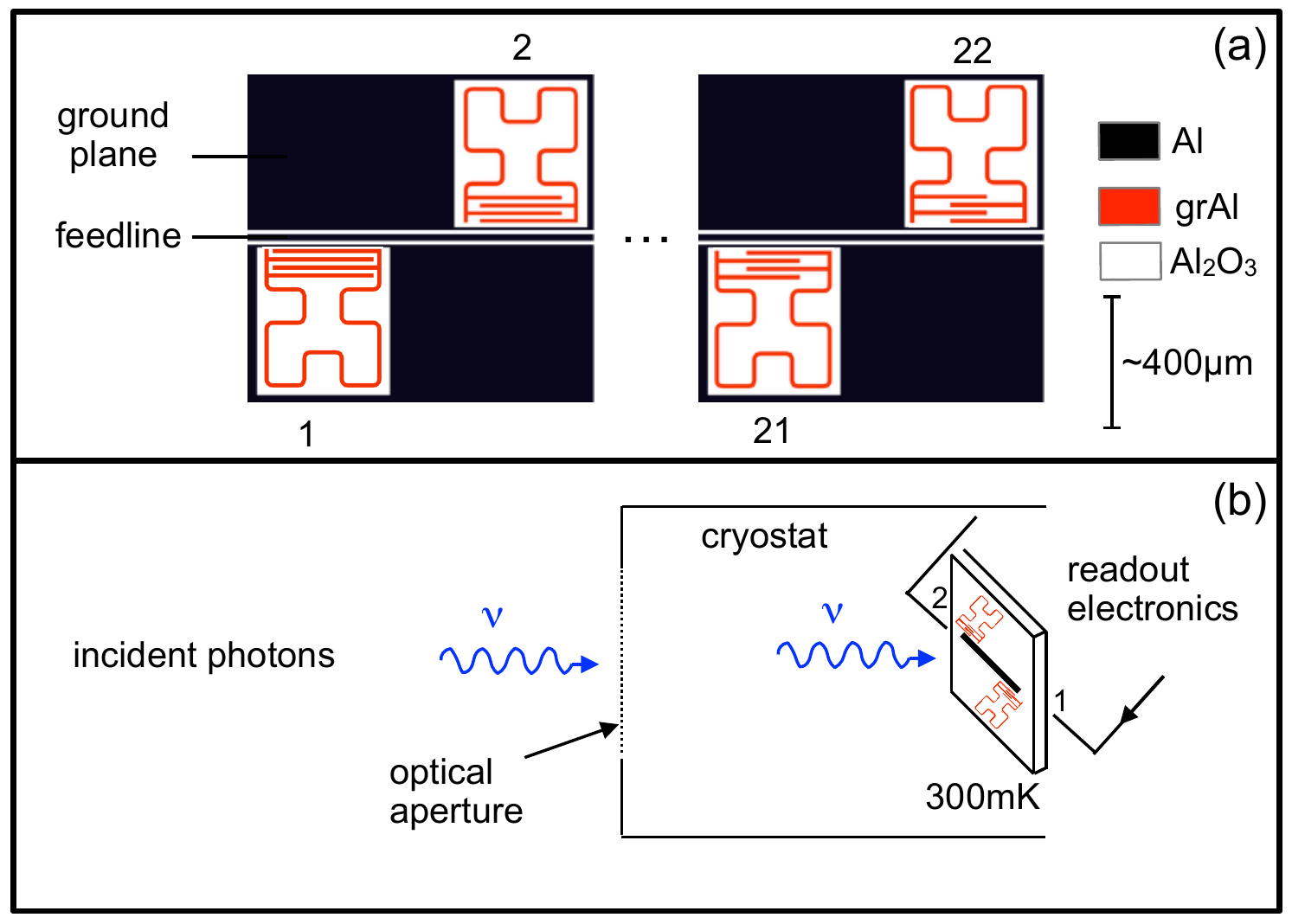}}
\caption{\textbf{Subgap kinetic inductance detectors (SKIDs) design and experimental setup.} (a) The SKIDs, in red, are planar resonators made out of superconducting granular aluminum (grAl). They consist in a Hilbert fractal inductor and a capacitive element that is adjusted to tune the resonant frequency. (b)  Photons illuminate the resonators through optical apertures and filters of a dilution refrigerator operating at 300~mK. The resonance frequency shifts are monitored through the transmission coefficient $S_{21}$ of the feedline.}
\label{fig_setup}
\end{center}
\end{figure}

Figure~\ref{fig_setup} shows the design of a SKID array (top panel) and a schematic representation of the experimental setup. The SKID array consists of 22 LC~resonators coupled to a feedline in a coplanar waveguide (CPW) configuration.  The inductor $L$, the radiation sensitive element, is identical for all the resonators whereas the finger lengths of the capacitors $C$ are adjusted to tune the resonant frequency $f_1=1/(2\pi\sqrt{LC})$ and achieve frequency multiplexing. The array is deposited on a 330-$\mu$m-thick sapphire substrate. The feedline and the ground plane are made of a 20-nm-thick aluminum layer with a normal state resistivity of $\sim$2~$\mu\Omega$cm, a critical temperature of approximately 1.4~K, and a superconducting spectroscopic gap $2\Delta/h\sim$100~GHz~\cite{catalano_bi-layer_2015}. The resonators are made of a 20~nm thick granular aluminum~\cite{cohen_superconductivity_1968,parmenter_isospin_1967,deutscher_transition_1973,dynes_metal-insulator_1981,pracht_enhanced_2016} layer with a normal state resistivity of approximately 900~$\mu\Omega$cm, a critical temperature of about 2~K, and a superconducting spectroscopic gap $2\Delta/h\sim$180~GHz. The array is cooled to 300~mK in an dilution refrigerator with optical access. Photons illuminate the resonators from outside the cryostat through a series of optical filters (see footnote for more details on the optical  filter stack~\cite{note}).
The filtering configuration ensured that only photons with frequency $\nu<$110~GHz reach the resonators.  Nineteen out of twenty-two  SKIDs are functional, with fundamental resonance frequencies ranging from 2.86~GHz to 3.79~GHz. 

\section{Results and discussion}

A standard protocol to quantify the efficiency of a photon detector is to evaluate its noise equivalent power  (NEP) ~\cite{catalano_bi-layer_2015, dupre_tunable_2017, catalano_sensitivity_2020, valenti_interplay_2019}, which corresponds to the signal power producing a signal-to-noise ratio of one in  a 1-Hz output bandwidth. NEP is defined in this case as
\begin{eqnarray}
\label{eq_NEP}
\textrm{NEP}=\frac{\Delta W_{\textrm{opt}}S_f}{\Delta f_1}
\end{eqnarray}
where $W_{\textrm{opt}}$ is the optical load power per detector, $\Delta f_1$ is the frequency shift of the resonance generated by the change of the optical load $\Delta W_{\textrm{opt}}$ (the power to be detected), and $S_f$ is the noise spectral density (NSD). The smaller the NEP, the more sensitive the detector. 

\begin{figure}
\begin{center}
\resizebox{8cm}{!}{\includegraphics{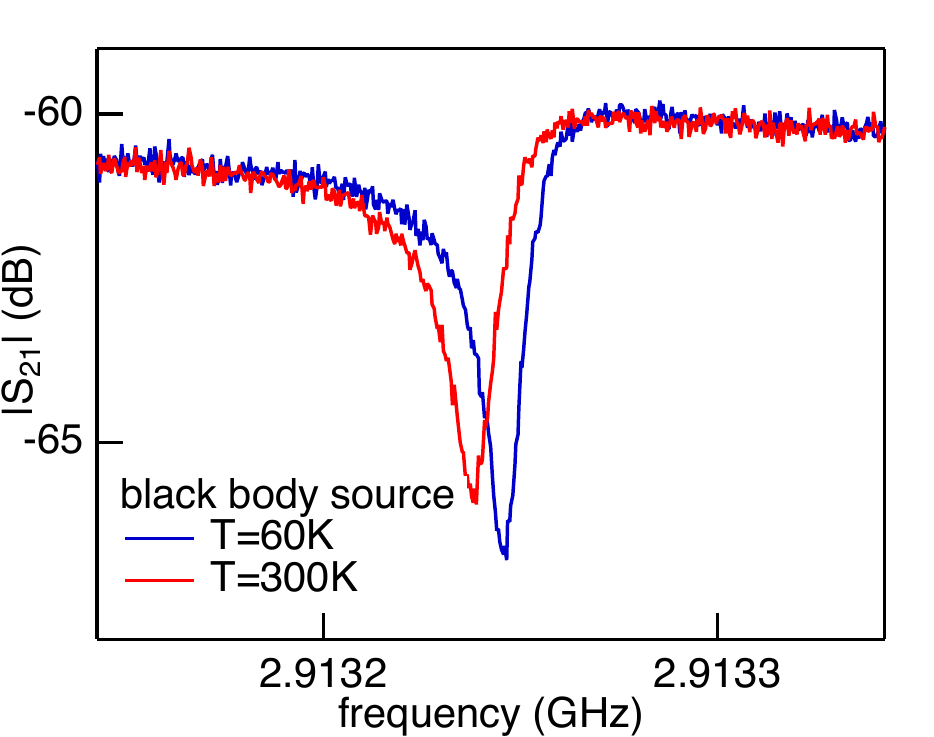}}
\caption{\textbf{Photon detection demonstration for SKID \#2.} The resonant frequency shifts down by 8~kHz when the temperature of a black body source illuminating the detector is increased from 60~K to 300~K. The black body spectrum is filtered by a low pass with a cutoff at 110~GHz.}
\label{fig_VNA}
\end{center}
\end{figure}

Figure~\ref{fig_VNA} displays the frequency shift of the resonance due to a change of the optical load. The figure shows the magnitude of the transmission coefficient $|S_{21}|$ of the feedline for the best SKID (\#2) for two different incident optical powers. The resonance frequency shifts by almost 8~kHz when the temperature of the black body source is changed from 300~K to 60~K. The average frequency shift of the array is 8.5~kHz, with a standard deviation of 1.4~kHz.

The evaluation of the change of the optical load per detector,  $\Delta W_{\textrm{opt}}$, is done using a three-dimensional ray-tracing software where the inputs are the spectral luminescence of the source (a black body source at 300~K  and 60~K), the geometry of the cryostat (the apertures, the distances, the lenses curvatures and materials), the size of the detector ($\sim400\times 400~\mu$m$^2$) and the spectral response of the detector. The spectral response of the detector can be overestimated by taking into account all the frequencies up to the low-pass filter cutoff frequency at 110~GHz, leading to $\Delta W_{\textrm{opt}}^{0-110~GHz}\sim 0.2$~pW. The spectral response can also be measured, as presented in the following, leading to a smaller and more realistic estimation of the change of the optical load $\Delta W_{\textrm{opt}}^{82-92~GHz}\sim0.05$~pW. This power corresponds to imaging at 85~GHz.

\begin{figure}
\begin{center}
\resizebox{8cm}{!}{\includegraphics{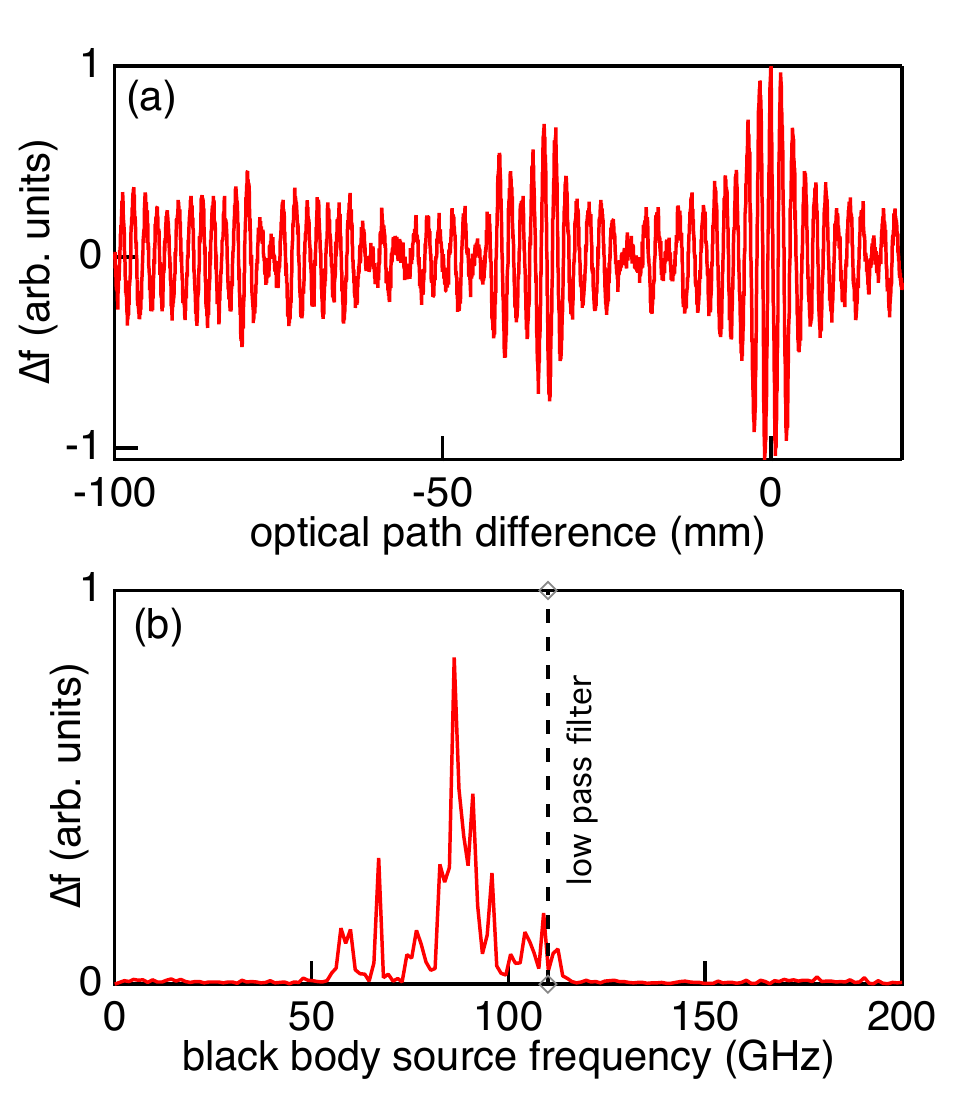}}
\caption{\textbf{Spectral response of SKID \#2.} (a) Frequency shift as a function of the optical path difference of a Martin Puplett spectrometer. (b) Frequency shift as a function of  incident photon frequency. Fourier transform of  the top panel response.}
\label{fig_FFT}
\end{center}
\end{figure}

Figure~\ref{fig_FFT} presents measurement of the spectral response of SKID \#2. The top panel shows the actual measurement: the frequency shift as a function of the optical path difference of a Martin Puplett spectrometer~\cite{martin_polarised_1970}. The Martin Puplett spectrometer is a Fourier transform spectrometer with a beamsplitter that consists of a grid of gold wires especially adapted for measurements in the 50~GHz~-~3~THz range~\cite{martin_polarised_1970,maleeva_circuit_2018}. The radiation source is a broadband and is split into two beams and recombined with an optical path difference that is varied thanks to a moving mirror. The intensity as a function of the optical path difference, the interferogram, is the sum of interference fringes of different wavelengths. The Fourier transform of the interferogram, the spectrum, is the intensity as a function of the incoming light wavelength. For our measurements, we proceed to a lock-in detection by modulating the incoming broadband source between a black body source at nitrogen temperature (77~K) and one at room temperature. The interferogram of the SKID \#2 (top panel) is almost a pure sinusoid with a few beating frequencies. The corresponding spectrum (the bottom panel) shows a maximum absorption at 86~GHz. The spectral response of the other SKIDs is similar: almost pure sinusoidal interferograms and spectra with a dominating absorption between 80 and 90~GHz. The SKIDs made out of granular aluminum with a normal state resistivity of approximatively 900~$\mu\Omega$.cm are thus intrinsically selecting the 80-90 GHz frequency band, when used with a low-pass filter of 110~GHz to suppress the above gap (standard KID) optical response.  

\begin{figure}
\begin{center}
\resizebox{8cm}{!}{\includegraphics{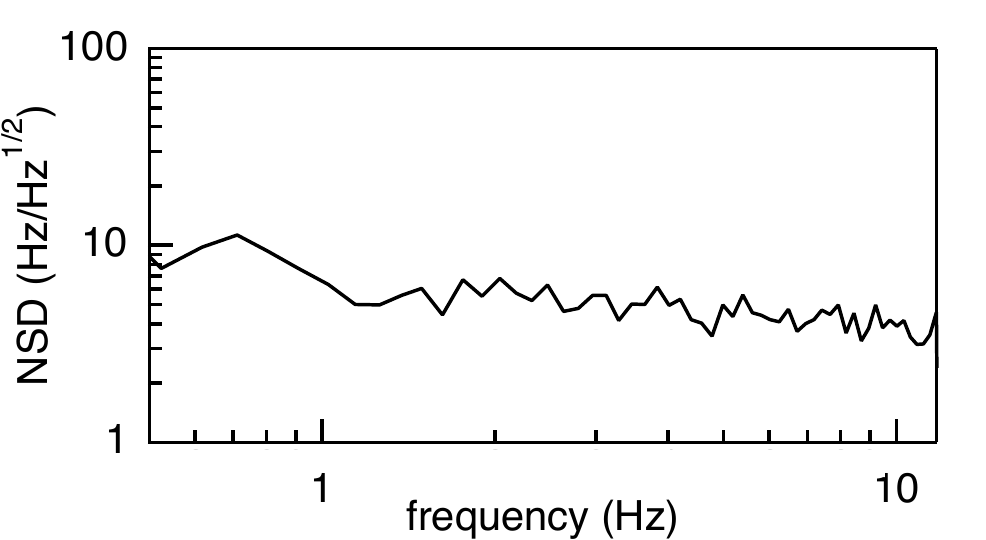}}
\caption{\textbf{Noise spectral density (NSD) of  SKID \#2.} }
\label{fig_bruit}
\end{center}
\end{figure}

Figure~\ref{fig_bruit} shows the NSD of SKID \#2. Under a constant optical load, the variation of the resonance frequency of each SKID is recorded over time and a Fourier transform is applied to obtain the NSD. At 10Hz, the NSD of SKID \#2 is  $N_f\sim$~4~Hz/Hz$^{0.5}$. At 10~Hz, the average NSD of the array is 28~Hz/Hz$^{0.5}$ with a standard deviation of 17~Hz/Hz$^{0.5}$. The uniformity of the array has to be improved. In particular, the response seems homogenous, but the noise is very variable. This last point needs to be understood. Optimization of the readout power for each individual SKID may help.

\begin{figure}
\begin{center}
\resizebox{8cm}{!}{\includegraphics{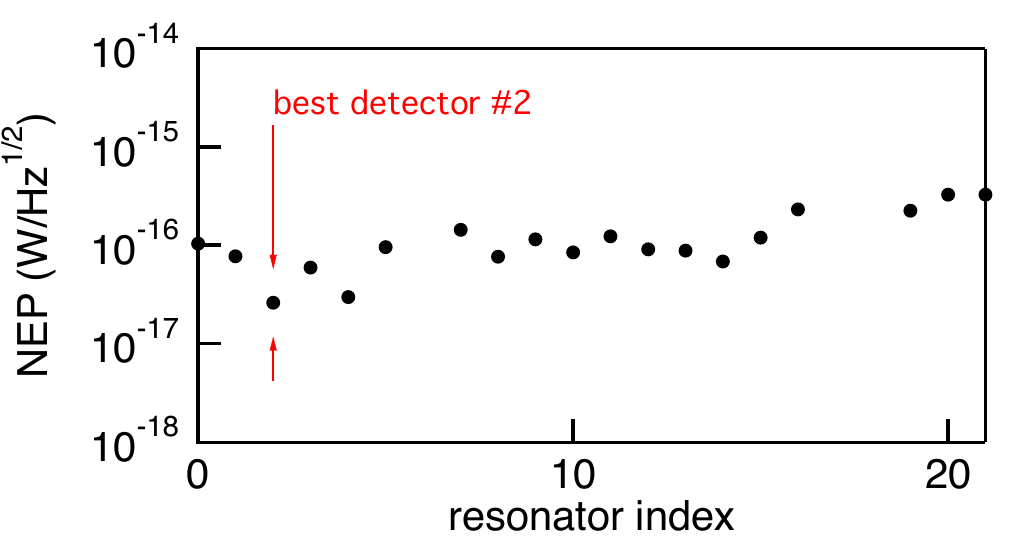}}
\caption{\textbf{Noise equivalent power of a SKIDs array for an 80-90~GHz frequency range.}  The SKIDs are identified by their index from 0 to 21, see figure~1a (19 out of 22 are functional resonators).}
\label{fig_NEP}
\end{center}
\end{figure}

Figure~\ref{fig_NEP} presents the evaluation of the noise equivalent power of our detectors with equation~(\ref{eq_NEP}), using the smallest value of the optical load variation.  For the best detector, SKID \#2, we obtain a NEP of $2.6\times10^{-17}$~W/Hz$^{0.5}$, comparable to the result obtained on KIDs made of a Ti-Al bilayer operating at 100~mK for a 80-120 GHz frequency band~\cite{catalano_bi-layer_2015}. The average NEP of the SKID array is $1.3\times10^{-16}$~W/Hz$^{0.5}$,  for an operating temperature of 300~mK and a 80-90 GHz frequency band.

\section{Conclusion}
The presented grAl SKIDs are potentially of interest for imaging applications in the 80-90 GHz band. Indeed, the upscaling of our array to a larger one is technically possible. The readout electronics are already  available to address up to 400 detectors per feedline. In addition, the base temperature requirement of 300~mK, instead of 100~mK, simplifies the cryogenic setup, although the need for sub-Kelvin temperatures still implies the use of relatively complex instrumentation. The intrinsic frequency selectivity of these detectors may be used as a natural band-pass filter. However, we still have not implemented a mechanism to tune the absorption band. Further investigation may increase the required minimum operating temperature 
by unveiling other superconducting materials with collective subgap modes and higher critical temperatures. Moreover, THz spectroscopy on chip may be envisioned if  a solution is found to tune the frequencies of the subgap modes, for instance by employing magnetic fields~\cite{borisov_superconducting_2020}, or dc current bias.

\section*{Acknowledgments}
We acknowledge the contributions of the Cryogenics and Electronics groups at Institut N\'eel and LPSC. 
This work is partially supported by the French National Research Agency through Grant No. ANR-16-CE30-0019 ELODIS2 and by the LabEx FOCUS through Grant No. ANR-11-LABX-0013.
F.L-B acknowledges financial support from the CNRS  through a grant  \textit{D\'efi Instrumentation aux limites 2018}.
F.V., N.M., L.G. and I.M.P acknowledge the Alexander von Humboldt Foundation in the framework of a Sofja Kovalevskaja award endowed by the German Federal Ministry of Education and Research. N.M. acknowledges the support of RFBR by grant 19-32-60064.

\end{document}